\documentclass[sigconf]{acmart}
\usepackage{listings}
\usepackage{sansmathfonts}

\begin{document}

\title{Efficient textual representation of structure}
\author{Brenton Chapin}


\begin{abstract}
This paper attempts a more formal approach to the legibility of text based
programming languages, presenting, with proof, minimum possible ways
of representing structure in text interleaved with information.
This presumes that a minimalist approach is best for purposes of
human readability, data storage and transmission,
and machine evaluation.

Several proposals are given for improving the expression of
interleaved hierarchical structure.  For instance, a single colon can
replace a pair of brackets, and bracket types do not need to be repeated
in both opening and closing symbols or words.  Historic and customary uses
of punctuation symbols guided the chosen form and nature of the improvements.
\end{abstract}

\maketitle

\section{Introduction}
Information is almost always more useful when organized, and structure is
key to that.  Therefore efficient and clear
representation of structure is of paramount importance.  Structured
programming languages are only one use of structure to organize
one kind of information, source code, but they make such unprecedentedly
elaborate use of structure that they have exposed deficiencies in our methods
of expression.

The languages for programming and math
developed in an evolutionary manner, with much borrowing from earlier work,
and the reasons for various decisions became buried in custom and history.
Studies on choices of characters and legibility have been patchy,
with some questions overlooked because they were
thought relatively unimportant.  Pioneers of programming languages
hurriedly made expedient adaptations of existing notations for similar
problems.  Most heavily borrowed was mathematical notation.

With many important questions needing settling with the creation of the
first programming languages, issues of symbology were mostly passed over
as unimportant and arbitrary.  As Bob Bemer noted,
``much documentation is lost, and it was characteristic of the times that
nobody seemed to think character sets a very important feature of
computers~\cite{bemer:18}.''
There are many studies on the readability and other properties of
various fonts and color combinations, but when discussed in relation to
source code, the term ``readability'' refers more to comprehensibility
~\cite{buse+weimer:2006}. 

Punctuation is the class of symbol most closely associated with showy,
interleaved structure.  Positioning is the other major method used to indicate
structure, and among the other intended purposes, ``control characters''
attempted to provide ways to position text.  Currently, Python is the most
popular programming language that relies on text positioning rather than
punctuation to indicate structure.  Visual programming goes further yet,
replacing textual indicators of structure and flow with graphical ones.

The programming language wars are still hot today, with new languages
emerging and gaining followers.  One cause of the passionate debates
is the tendency of language designers to resort to an evangelical
approach to justify their choices of design elements for which they have
little compelling technical reason.  Sometimes the 
designers make overlarge and unsubstantiated claims~\cite{markstrum:2004}.
For many programming languages, one of the defining features is the
choice and usage of symbols.  These choices are not modifiable by the
programmers, so that if such changes are desired, a whole new
programming language may need to be created, another factor in the very
proliferation of programming languages that the ALGOL designers were hoping
to avoid.

  The ideas in this paper
aim at the foundation, the symbolic representation of the structure.
Structure is chosen as the crucial concept
that must be addressed to improve representation.  Minimalism is
the chosen guide.

Too much minimalism is certainly possible, for instance by
expecting people to work with information that has been minimized by
data compression techniques which transform the data into a very compact but
human unreadable form.  The minification techniques of removing unnecessary
whitespace and shortening variable names is another example.
The target of these minimization efforts is the representation of the 
structure of the source code, not the source code itself.
Further, this is not about rewriting
and rearranging to place information in more efficient structures, this is
about making the representation more efficient regardless of the structure
chosen.  Punctuation has always
been minimal, using smaller and less obtrusive symbols than
those used to represent the letters of a language, and the syntax of
structural elements follows that pattern.

Some more items to note are splits between textual representations used
for source code, versus those used for markup, as in HTML, and data
organization, as in XML and YAML.  Within programming languages, 
there is the dot (or arrow) notation of Object Oriented Programming and
the completely different notations for Structured Programming,
such as the curly braces.  Yet those splits seem artificial, as hierarchical
structure is used in all.  Many programming languages
are needlessly poor at expressing data.  Several of the
improvements in C++11 and C++14 touch on this issue, allowing more flexible
constructions of constants and initializations of arrays and objects.
One intent of JSON is to bridge this divide.

Also, these are all interleaved formats, meaning
the symbols that denote the structure are interleaved with the symbols
of the data.  Goals in data storage are minimal size, and fast access.
An obvious and common method to achieve both is to exclude all complex
structure from the data, using an external representation.  The disadvantage is
that they require some connection, often expressed in fixed sizes and padding,
which can end up using more space than an interleaved method.  Over the
years, the attention paid to brevity has varied with the cost and
availability of storage.

By 1963, the American Standard Code for Information Interchange (ASCII) 
was set to a fixed size of 7 bits, though
at least two variable codes, Morse code (1844) and Huffman coding
(1952) existed at the time.

One of the goals of XML was human readability.  Minimalism was
thought orthogonal or possibly even antithetical
to the goal of human readability, and the resulting language ironically
suffers from excessive verbosity that obscures essentials, rendering
the result less human readable.  XML and COBOL show that a negative
attitude towards minimalism ("10. Terseness in XML markup is of minimal
importance.~\cite{xml:1996}"), that regarding minimalism as unrelated or
even an impediment to comprehension, is not correct.  Minimalism is also
central to Information Theory, in which it was demonstrated that the crude
redundancy of repeating
information over and over, is very wasteful and poor at preserving the
fidelity of data against errors in the transmission.  If repetition is a poor
means of ensuring the fidelity of data, perhaps it is also a
poor means of representing structure in ways easy for humans to read.  Another
demonstration of the
limited usefulness of repetition is the FAT file system, which despite
allocating room for a copy of the directories and file names, is actually
one of the most fragile and easily corrupted file systems currently in use.

Of particular note is the C programming language.  So many programming
languages adopted the C syntax that they have been tagged with a moniker
of their own, the
``curly-brace'' languages.  Perhaps one of the reasons curly brace syntax
eclipsed Pascal and ALGOL is the use of a single character each, rather than
the words BEGIN and END, to delimit blocks.
The designers of C did not restrict themselves to curly braces only,
they also used square brackets, parentheses, and even angle brackets,
for array indexing, lists of function parameters, and macros respectively.
Why that choice of symbol assignment?  Why not use parentheses
for all, and rely on context or some other means to distinguish between
a parameter list and a block of code?  If there is any doubt that it is
possible to use only parentheses, the LISP programming language is proof.
Or, why not copy FORTRAN in the use of parentheses for array indices?
One kind of answer is that in C these different kinds of brackets serve as
sigils, to distinguish between identifiers for functions, arrays,
and variables.  But that only begs the question of why have sigils?
And it still does not answer why any particular symbol was chosen for
a particular use.

\section{History}

For answers, one must dig into the history of computation and mathematics.
In the case of C, the chain of preceding languages is roughly B, BCPL
(Basic CPL), CPL (Combined Programming Language), and finally
ALGOL (Algorithmic Language).
The paper on ALGOL 58~\cite{algol:1958} says of the choice to use square
brackets to delimit array indices, only that ``subscripted variables'' (the
term used in ALGOL for what today we call an array variable, or simply
an array),
``designate quantities which are components of multidimensional arrays''
and that ``The complete list of subscripts is enclosed in the subscript
brackets \texttt{[]}.''  But why did they pick square brackets?
FORTRAN, the oldest programming language to achieve wide acceptance,
uses parentheses, not square brackets.

For that matter, why use any bracket at all?  No one says.
It seems likely that they would rather have used actual subscripted
text, just like in mathematical notation, but early computers
could not do it.  Square brackets was a notational device to
indicate subscription without actually presenting the text so.
Apart from computer limitations, a big problem with subscripting is that the
notation doesn't nest well, at 3 or more levels becoming too small for the
human eye to read.  One can surmise from the
use of the term ``subscript'' that this was another borrowing, from
linear algebra in which a matrix is denoted with square brackets.  And indeed
the original name of ALGOL 58, is International Algebraic Language.
The only deviation in the use of square brackets for array indexes 
from ALGOL to C was BCPL, which among the many
simplifications of CPL it introduced, attempted to
repurpose square brackets for code blocks, using only pointer
arithmetic to access array elements~\cite{martin:1969}.

ASCII codified the glyphs used for nearly all programming languages.
A notable exception is APL, which makes use of mathematical symbols,
mainly from Set Theory and Vector calculus, that were not put in
ASCII~\cite{iverson:1962}.
Unlike EBCDIC, ASCII at least organized the alphabet into a contiguous block.
But the exact set of punctuation symbols is unclear, ranging from all symbols
that are not letters, numbers, and control characters, to only those used
to clarify the structure and meaning of sentences and text.  There are no
formal, ordered, centuries old lists of punctuation symbols.

The ASCII ordering and choice of punctuation is derived from the QWERTY
keyboard layout, which dates to the late 19th century.  The notion that
QWERTY was deliberately arranged to slow typists down is a popular but wrong
myth~\cite{yasuoka+yasuoka:2011}.  Morse Code and many other factors were
considered, and over the years small changes have been made to accommodate
new uses.  For instance, ``shift-2'' is the double quote
mark on many older keyboards, but today is `@' on most keyboards.

We could go further back, and ask why mathematical notation uses
parentheses for functions, and square brackets for matrices.  Why is
$y=f(x)$ the customary, canonical expression for a function,
and why in particular the use of parentheses to bracket the independent
variable $x$?
In A History of Mathematical Notations~\cite{cajori:1928},
Cajori credits Euler
(1707-1783) with the first use of parentheses to bracket
the variable of a function, in a 1734 paper.  That paper is E44~\cite{euler:44}
in the numbering scheme created to refer to Euler's works.
However, examining E44 and several others of Euler's papers, one finds
no such use of parentheses, and the exact phrase and
formula Cajori quoted is not present.  Euler uses parentheses to group
parts of equations, but not to separate function names and variables. 
Euler's notation is $y=fx$, and 
it is up to the reader to understand that $x$ and $y$ are variables,
and $f$ is a function.

Note also the choice of the letter $f$ because it is the first
letter of the word ``function'', a custom followed in many places,
such as the decision in FORTRAN to use the first letter of a variable name
to indicate
integer (name begins with ``I'' for integer, through ``Z'') or floating point
(name begins with ``A'' through ``H'').  This desire to match functionality
to the first letter of an appropriate term was taken to extremes, so that
more than one early game employed a lucky placement of keys on the QWERTY
keyboard,'W', 'E', 'S', plus '3', to refer to west, east, south, and
north respectively.

By 1837, in a major work on Number Theory which is regarded as also an
important paper on the modern definition of a function, Dirichlet (1805-1859)
used parentheses around the independent variable~\cite{dirichlet:1837}.
But why did mathematicians pick those symbols, that format?
They too engaged in expedience, adopting 
the idea of parentheses from still earlier scholars.
Mathematical notation has a long evolutionary history, and while fascinating,
the main point here is that many choices of symbols and syntax were made
long before any possible use in programming languages was conceived.  While
1837 is also the year that Babbage proposed the Analytical Engine,
arguably the first computer, functioning computation machinery would not
be built until many years later.  Therefore symbols and syntax certainly
could not have been chosen based on experiences in computer programming.

That was about as far as the early pioneers went in exploring questions
of how best to symbolize code and data.  None of the terms and areas
of study, not semiotics, symbology, linguistics, grammar, lexicology,
punctuation, readability, typography, legibility, notation,
expressiveness, or rubrication, quite
address these questions.
Studies of notation and syntax get the closest, but even there syntax
is confined to issues of context.

Most programming languages use a hierarchical structure to
organize code.  Possibly the earliest and simplest formally specified
language for expressing hierarchy is Dyck Language.
Object Oriented Programming and Functional Programming did not abandon this
fundamental organization, they only added to it.
Declarative programming, as represented in Prolog and SQL, at first glance
seems not to need much structure.  A point of confusion is order vs structure
vs hierarchy.
Declarative programming needs structure, but not order and not necessarily
hierarchy.
Hierarchic structure, of programs and data, can be more efficiently represented
with several changes.

The advent of markup languages revived interest in hierarchical data storage,
which was introduced in the 1960s, before the relational database model.
No longer were interleaving structural symbols just for programs,
they were harnessed to organize data.
Traditionally, data has been organized into fixed size elements
so that no symbols need be reserved for explicit denotation of structure,
and, even more importantly, so that random access is quick, taking
$O(1)$
time to retrieve any one element.
This is also true of the pre-computer era, which used tables extensively,
carefully lining up columns to aid the human eye.
Where one-size-fits-all is inadequate, the expedient method
used is to have a small fixed size field to hold a value for the size
of a variable length field.  Packet networking is an example of this
organization of data.  The roughly analogous method in writing is the
technique of employing any of a variety of superscripted symbols such
as an asterisk, *, or a dagger, $\dagger$, to indicate there is a footnote.

XML and HTML are the most well known of these markup
languages, and like programming languages, their history is also
evolutionary.  Both trace back to Standard Generalized Markup Language
(SGML) which was standardized in 1986, predating the World Wide Web.
Like so many other decisions in languages, the creators of the Web seized
upon SGML out of expediency. SGML in turn descends from GML, an IBM effort
to manage technical documentation and data, based upon ideas first
articulated circa 1966~\cite{goldfarb:1997}.

But as many have complained over the years, these markup languages have
undesirable features, and among the biggest is extreme verbosity.
The rules they force upon users, to come closer to the goal
of ``human readability'', often have the opposite effect.
On the scales of minimalism, XML and relatives are extremely
poor because their representations are highly redundant.
Not only must brackets be balanced in ``proper'' HTML and XML, but
the matching tags must repeat the tag name.
Why did the designers do it?
Ironically, those rules have done much to add clutter
and thereby reduce the human readability that was their intended goal.
YAML (YAML Ain't Markup Language) was motivated in part by recognition that
XML is burdened with design constraints that have little purpose in
data serialization~\cite{yaml:2004}.
Lightweight markup languages such as Markdown
are an acknowledgment that the human readability of HTML could
be better.

Most popular programming languages are poor at expressing data.
Here are some examples to illustrate this.  A list of the first 10
chemical elements can be encoded in a JavaScript array like this:

\begin{lstlisting}
const CE=["?","H","He","Li","Be","B",
          "C","N","O","F","Ne"];
\end{lstlisting}

A simple trick yields a much cleaner representation:

\begin{lstlisting}
const CE="? H He Li Be B C N O F Ne"
         .split(" ");
\end{lstlisting}

But this is the very sort of trick that makes programming
needlessly difficult for professional programmers unfamiliar with the
arcana of a particular language.

One problem is that the default, unquoted
meaning of an alphanumeric sequence is to treat it as the name of a variable.
The double quote mark changes the mode, but that mode has no support for
structural elements, so only a simple string can be encoded.  The programmer
is forced to change modes over and over, entering string mode to give a
short string, leaving string mode to impart a tiny amount of structure,
then entering string mode again to give the next string.  Or the programmer
can use a clever trick such as the {\tt split} function, or 
create a function to parse a string into a complicated object, or
even employ a library such as YAML.

Another example, of a family tree, in Python:

\begin{lstlisting}
class tn:  # tn means "tree node"
  def __init__(self,name,child=None):
    if child==None: self.c=[]
    else: self.c = child
    self.n = name

familytree =
  [tn("grandmother",
    [tn("older uncle",
      [tn("oldest 1st cousin"),
       tn("2nd oldest 1st cousin")]),
     tn("father",
      [tn("older sister",
        [tn("niece"),
         tn("nephew")]),
       tn("you",
        [tn("son",
          [tn("granddaughter")]),
         tn("daughter",
          [tn("grandson")])]),
       tn("younger brother")]),
...
\end{lstlisting}

This terrible encoding is littered with alternating 
brackets of 2 kinds, as well as double quote marks and commas.
This shows that Python can be even worse than LISP,
for those who thought Python's use of indentation lead to clean code
in all cases, and that LISP had too many parentheses.  To get clean
looking code, the expert programmer resorts to using
functions to read a simple string (which may be a data file)
into a complicated object.  Employing
a data serialization library such as YAML,
is a common method of handling this issue.
Should it be the preferred method?  Shouldn't
programming languages be able to do better with their native syntax?
After all, native handling of regular expressions is what made Perl
popular.  Improvements in the representation of structure are applicable
both to coding and to data representation.

\section{Eliminating runs of brackets}

The first change addresses a problem most languages have,
but which is perhaps most obvious in LISP, and for which it
has been criticized in the ``backronym'' of Lots of Idiotic Spurious
Parentheses.  Often, brackets cluster,
as several structures all start or end simultaneously.  They can add to the
visual clutter without adding to the ease of comprehension.

There are many solutions to this problem, among them operator precedence,
and postfix notation, also known as Reverse Polish notation,
first conceived in 1924\cite{lukasiewicz:1931}.  A limitation of these
Polish notations is that to make brackets unnecessary, the number of
operands must be fixed, an inflexibility that is insufficiently general
for the structures used in programming. 

A popular short cut is use of context and knowledge about the permitted or
sensible content
of subtrees.  For instance, in HTML the paragraph indicator, {\tt <p>},
cannot be nested.  This is often used to omit the matching closing 
bracket, {\tt </p>}, when the next structure is another paragraph, or
something else that cannot be inside a paragraph, such as a header.
Such omissions are not officially sanctioned in HTML, but are so popular
that web browsers had to support them anyway.  Obvious
problems with this approach are that knowledge of every exception to
the rules for indicating the nesting may be very large, and may change.

The approach taken in Perl 6 is to allow all kinds of shortcuts that
do not greatly complicate the parser.  Compared to Perl 5, some brackets are
no longer required.  In particular, the parentheses of the {\tt if} and
{\tt for}
statements are optional~\cite{perl6synopis4:2004}.  Effectively, this change
is a recognition that {\tt if} and {\tt for} are enough by themselves to
indicate structure, that they are in fact now part of the set of symbols
used to denote structure.

One could employ 2 sets of brackets, perhaps {\tt ()}  and {\tt []}, in a
scheme in which a closing bracket closes its matching
opening bracket, and all the open brackets of the other kind in between.
For example, 
{\tt [a [b]]} becomes {\tt [a (b]}, {\tt [d [e [f]]]} becomes {\tt (d [e [f)}.
This idea can work in the other direction.
{\tt [[g] h]} becomes {\tt [g) h]}.  It even works in both directions at once,
with {\tt ((j)(k))} becoming {\tt [j)(k]}.  However, the best this idea can
do for {\tt ((m))} is {\tt [(m]}.

An issue is that 2 more symbols are needed.
We can employ only one more symbol, eliminating only one of the excess
opening or closing brackets, and still clean up most of clutter.
Call a 3 symbol system that eliminates excess closing brackets a ``closing 3'',
and a 3 symbol system that eliminates excess opening brackets an ``opening 3''.
Using colon, {\tt :}, for this 3rd symbol in a closing 3 system, because that
approximately
matches the traditional use of the colon in written natural languages,
changes {\tt (a (b))} into {\tt (a : b)}.  {\tt ((m))} becomes {\tt (:m)}, {\tt (((n)))}
becomes {\tt (::n)}, and {\tt ((j)(k))} becomes {\tt ((j):k)}.
Additionally, the brackets are still balanced, with equal numbers of opening
and closing brackets in all the systems.

For a slightly larger example, consider this Ackermann function,
from the classic textbook Structure and Interpretation of Computer
Programs, exercise 1.8~\cite{abelsonsussman}:

\begin{lstlisting}
(define (A x y)
  (cond ((= y 0) 0)
        ((= x 0) (* 2 y))
        ((= y 1) 2)
        (else (A (- x 1)
                 (A x (- y 1))))))
\end{lstlisting}

Employing a closing 3 system as suggested above, gives this:

\begin{lstlisting}
(define (A x y)
  :cond ((= y 0) 0)
        ((= x 0) :* 2 y)
        ((= y 1) 2)
        :else :A (- x 1)
                 :A x :- y 1)
\end{lstlisting}

6 colons have replaced 6 opening brackets.  The 6 matching closing
brackets have been removed.
Indeed, there is never a need for multiple adjacent closing brackets,
as proven next.

\begin{theorem}
Given a sequence $S$ of arbitrary symbols over an alphabet $A$
in which 2 symbols, an ``opening'' and a ``closing'' symbol, are reserved
to denote hierarchy in a format that interleaves data and structure,
and $S$ is properly balanced, the hierarchy can always
be represented in a system with 3 reserved symbols in which there are no
runs (sequences of length 2 or greater) of the closing symbol.
\end{theorem}

\begin{proof}
WLOG, let `{\tt(}' and `{\tt)}', the parentheses,
represent the opening and closing symbols in both
systems, and let `$:$', the colon, represent the 3rd symbol in the 3 symbol
system.
To allow elimination of all runs of 2 or more closing symbols, assign `$:$'
the same meaning as `{\tt(}', the opening of a subtree, except that the
matching closing symbol for `$:$' is an already necessary `{\tt)}' that matches
an existing `$($' which precedes the '$:$'.

Then, instances of the sequence ``{\tt( $s_1$ ( $s_2$ ))}'' in which $s_1$
and $s_2$ are arbitrary sequences which may include balanced occurrences
of `{\tt(}' and `{\tt)}' and `$:$', may be replaced with
``{\tt( $s_1$ : $s_2$ )}''.

The replacement symbols are sufficient to represent all the relationships.
The symbols still indicate that $s_1$ is the parent of $s_2$, preserve
all relationships $s_1$ and $s_2$ have with all other sequences before and
after because none of them need change and no additional context is
needed, and preserve all relationships contained within and between $s_1$
and $s_2$ also because none of them change, nor add any contextual
dependencies.

This replacement can be applied repeatedly, to reduce any number
of adjacent closing brackets to 1 closing bracket.  Each replacement
preserves the property of balance for all remaining parentheses,
as exactly one pair of matched parentheses is replaced with a single colon.
\end{proof}

The corollary that no runs of the opening bracket are needed
in an opening 3 system, is obvious.

A pushdown automaton can easily transform from an opening 3 to a 2,
or from a 2 to a closing 3, if the data is processed in reverse order.
Of course, a pushdown automaton can easily reverse a string.  In practice,
the C++ style of comment delimited by a 2 slashes takes more work to detect
from back to front.  Nor can the start of a C style comment be
simply determined working from back to front, because ``/*'' can be within a
comment.

A natural question is why not use a 4 symbol system, as originally outlined
above with the 2
sets of bracket symbols, and eliminate all runs of opening and closing
brackets?  Simply put, the additional savings is not significant, as can be
seen in that it is no help at all on the examples
of {\tt ((m))} and {\tt (((n)))}.

As to why, it is not possible to employ any 
finite set of symbols to represent infinitely many numbers
with just 1 symbol each, no matter what form the representation takes.
If the representation
takes the form of $n$ opening brackets followed by $n$ closing brackets,
all of one kind of bracket can be collapsed, because $n$ is preserved
in the other.  If both are collapsed, then $n$ must be represented some other
way.  That is why the idea of using 2 sets of brackets does not
work to reduce all runs of opening and of closing
brackets to 1.

Thus we see that the idea of replacing each run of closing brackets with a
single closing bracket is really the removal of a redundancy, the redundancy
of specifying the depth twice, first with opening brackets, then
with an equal number of closing brackets.  That redundancy is no
longer available to remove once one kind of bracket has been reduced.

The 3 symbol system need not be exclusive, can mix with 2 symbol usage
as in {\tt(a (b : c))}.  In practice, in coding it will likely be preferable
to use the 3rd symbol only for subtrees that are known in advance to be the
terminal child.  For other uses, such as
minification of JavaScript or JSON, may want to use the 3rd symbol everywhere
possible.

Removing the redundancies of the 2 symbol system can be of some
value in data compression.  Since the amount of information encoded is the
same, an ideal data compression algorithm should produce the same size
compressed output whether a 2 or a 3 symbol system is used.  In practice,
the output sizes vary, sometimes better for the 3 symbol system, and
sometimes worse.  To better test whether the more efficient representation
helps with data compression, can try a much larger example. 
Biologists have organized millions of species into a Tree of Life
~\cite{opentree:2017}, using
Newick format~\cite{newick:1986}, an interleaved hierarchical format.
Tests upon grafted\_solution.tre from version 9.1, the file with the highest
percentage of
interleaved structural symbols relative to data, containing 100,734
parentheses in 721,324 characters total, show an ``opening 3'' system
does reduce size even after compression.

\begin{center}
\begin{tabular}{lrr}
compression  & system \\
       & original 2 symbol  & opening 3  \\
none   & 721,324  & 690,077 \\
gzip   & 250,142  & 241,169 \\
bzip2  & 218,717  & 213,341 \\
xz     & 211,812  & 203,724 \\
\end{tabular}
\end{center}

A final note about whether to prefer an opening 3 or a closing 3 system.
The closing 3 is the better fit with our customs.  For instance, in curly
brace languages, the name of an array is given before the index of the
desired element.  It is {\tt arr[39]} not {\tt [39]arr}.  It is the same with
function names and parameters-- the name comes first.

\section{Universal bracket}

A sequence such as ``\verb|[x(y]z)|'' in which 2 different sets of brackets are
interwoven, is almost always an error, not valid in any
mainstream language.  An analogous sequence in HTML could be
``\verb|<b>x<i>y</b>z</i>|'', which is not valid, even though its
meaning can in this case make sense.  The HTML specification calls this
``misnesting''.
This invalidity is used in an ad hoc fashion to reduce some
of HTML's redundancy.
A common case is the closing of an outer structure that
implies an inner structure must also close, as in this example:
``\verb|<tr>x<td>y</tr>|''.  Some omissions require
knowledge that some structure is not allowed.  For instance,
``\verb|<p><p></p></p>|'' is not valid because the `p' element (p for
paragraph) can't be the direct child of another `p' element.  Therefore
``\verb|<p>x<p>|'' always implies a closing tag preceding the 2nd opening
tag: ``\verb|<p>x</p><p>|''.
This usage is acknowledged in HTML5, but still recommended against:
``..the closing tag is considered optional.  Never rely on this.
It might produce unexpected results and/or errors if you forget the end tag.''
~\cite{HTML5elements:2016}

A combination opening and closing tag in one, called a ``self-closing'' tag,
is meant for an empty element, and has been in XML from the start
~\cite{xml:1996}.
As of version 5, HTML has adopted a variation of this idea.
The XML self-closing tag requires a slash character immediately before the
closing angle bracket.  In HTML5, 15 element types were singled out as making
sense only as empty (void), and HTML does not use or permit the penultimate
slash character in those tags.

Another solution to some of HTML's verbosity is to omit the name from
the end tags, using only ``\verb|</>|'', which works fine since misnesting 
is not allowed or often sensible anyway.  SGML
has this feature in its SHORTTAG constructs, calling it
the empty end tag.  But HTML does not allow it.  This idea of
a universal closing bracket or, alternatively, a universal opening bracket,
can be employed in any language containing
2 or more sets of bracket symbols and in which interweaving is invalid.
It eliminates misnesting, as interweaving is no longer possible.
And it reduces the alphabet size.

If we choose the square bracket for the universal closing symbol,
then a sequence such as ``\verb|(x[y]z)|'' could become ``\verb|(x[y]z]|'',
and the closing parenthesis symbol would be unused,
and could be repurposed.  (Note that this change does not reduce the number
of closing brackets, there are still $2$ in the example.  It reduces the
required size of the alphabet.) 

There can still be a need for other closing characters, such as 
an ``unindent'' invisible control character.  The universal closing
bracket could still be used for that, but would want it to be invisible
in that case.

Converting back and forth between a representation
that uses closing brackets that match the opening brackets, and a
representation that uses only a universal closing bracket is easily done
with a pushdown automaton.  The type of the node is preserved in the
choice of opening
bracket, and having the type repeated with a matching closing bracket
is merely redundant.

Having established that a universal bracket symbol is workable,
several more questions naturally arise.  Does it make code easier to
understand, more human readable?  Many have expressed the sentiment that
requiring a closing tag to repeat the type given in the opening tag helps
prevent human mistakes, and is therefore good.  The issue is confused
by the practice of entirely omitting tags in specific situation.  With
a means of representing structure that is not so tiresomely redundant,
these ugly short cuts can be made unnecessary.

\section {Types for nodes}

Often the representational capability of a hierarchical
structure is enhanced by adding some means of
representing different kinds of children.  An example is the ``red--black
tree'' in which tree nodes have been assigned an additional property, a color.
This can be and is often done independently of the
structure, by means of an additional data item.  Another very popular
method is sigils in the form of different kinds of brackets.
ASCII has 3 sets of symbols meant solely for brackets:
the parenthesis, square bracket, and curly braces.  One more set, the
angle brackets, doubles as the mathematical symbols ``greater than''
and ``less than'', and for that reason was used gingerly.
Further sets can be contrived, for instance `$\backslash$' and `/', and for
that matter of course any two arbitrary characters could be chosen to
serve as brackets.  The obvious complaint
is that 4 sets is far too few.  Even if a dozen plus from Unicode
are added, it still isn't enough.

In any case, programming language designers used all the ASCII symbols
meant for brackets early on.  The curly brace languages employ curly
braces to denote blocks of code, square brackets to denote array indices,
and parentheses for parameter lists.  The dual purpose symbols for
angle brackets did not go unused as brackets, being employed in the
C preprocessor, and later in the markup languages SGML, XML, and HTML.

These SGML markup languages expanded the number of bracket types infinitely,
by allowing multiple character brackets. Although that solves the problems
caused by finite quantities of different bracket symbols, the means and
requirements chosen add greatly to the verbosity, a common criticism
often expressed in abuse of the rules rather than in words.  It is possible
that the desire for a visual match between the opening and closing bracket
led to the SGML requirement that both the opening and closing brackets
contain copies of the string employed to give them a type, despite the obvious
redundancy.

An efficient way is to designate one of the bracket sets as "typed",
the start of a multicharacter bracket.
That allows the other brackets to remain bare
to be used same as traditionally used in programming languages, and still
allows infinite bracket types.  Which symbol is best employed, and where
should it be positioned?  Between bracket and name, or after the name?
Or, should it be a combined symbol, a bracket that indicates a name is
to follow, since there is more than one kind of bracket available?
Possibly the most efficient use of existing symbols is to keep parentheses
as is, bare, and designate the square bracket or curly brace as the
indicator for a child structure with a type, with the name to follow,
similar to HTML.

Another method is to reserve a symbol to indicate a type name only,
no structure.  `\$' is often used similarly.

Whichever method is chosen to indicate the start of a type name,
how is the name to be ended?  The name could be just like variable
names in programming languages, with only letters and numbers (and the
underscore character) allowed in the name so that any other symbol, such
as the space character, automatically ends the name.  The method of
using a special symbol, as done with the closing angle bracket of HTML,
is also workable.

But the designers of HTML did not let tag names be only names.  They crammed
additional
structured information into ``attributes''.  An example is
``{\tt <ul class="vs1" id="item1"> content </ul>}''.
This information
could have been in the same system, for instance something
like ``{\tt<ul> <attr> class = "vs1" id = "item1" </attr>
content </ul>}'',
or even ``{\tt<ul>} {\tt<attr>} {\tt<nm>}{\tt class} {\tt </nm>}
{\tt<val>} {\tt vs1}  {\tt</val>}
{\tt<nm>} {\tt id}  {\tt</nm>} {\tt<val>} {\tt item1} {\tt</val>}
{\tt</attr>} {\tt content} {\tt</ul>}''.
The only purposes this alternate subsystem really serves is visual
distinction and less verbosity, though it's claimed to
maintain the distinction between data and metadata.  HTML has evolved
towards lighter use of attributes, moving much formatting information
from the tags to CSS, where it is also less redundant.

\section{Representing siblings and cousins}

The list is well known and has a long history.  Each item in a list can be
considered a sibling of each other item.  Traditionally, each item is on its
own line, or is separated by a comma.
LISP means ``LISt Processor'', and is built around the idea of making lists
the fundamental building block with which to organize both data and code.
Comma Separated Values (CSV) notation~\cite{RFC:4180} is a simple
data format based on one list with items separated by, of course, commas.
One of the most notorious departures from the use of commas is multidimensional
arrays in {\tt C}, in which the syntax to access an element at index $x,y$ is
not $[x,y]$, it is $[x][y]$.

The idea of separating items in a list with a single symbol (or word)
seems simple, but turns out to have several surprisingly
tricky issues.

Consider how to represent
a list in a language that does not have any symbol analogous to the comma,
Dyck Language interleaved with data.  How is the sibling
relationship expressed?
(First, note the convention is to place the parent before
the child, as in \verb"p(c)", although the opposite, \verb"(c)p" is
just as expressive.)  One way is to wrap brackets around each individual data
item.
Then the number of brackets needed to represent a relationship must be
increased by 1 for all depths, so that \verb"(a)(b)" means $a$ and $b$ are
siblings, and \verb"((c))((d))" means $c$ and $d$ are 1st cousins.
A 2x2 array would be {\tt ((p)(q))((r)(s))}.  Although it works,
it is far more verbose.  Additionally it spoils the abbreviation of allowing
siblings
to be separated by a child, as in \verb"a(e)b", which must instead be
\verb"(a(e))(b)".
So, a better way is to always separate siblings with a child, using
a null child if the older sibling has no children,
as in \verb"a()b".  Then a 2x2 array can be represented with
{\tt (p()q)(r()s)}.

Expanding to cousins is still a problem.  With the addition of the comma
as a sibling separator,
{\tt (p()q)(r()s)} becomes
{\tt (p,q)(r,s)}.  The sequence still has a ``{\tt )(}'', which the comma
does not help reduce.
An obvious extension is to
introduce another symbol, say semicolon, to separate 1st cousins.
Then the sequence can become {\tt (p,q;r,s)}.

What to do for 2nd cousins?  Just add brackets to the semicolon,
as in {\tt );(}?
Or employ yet another symbol to replace {\tt ))((}?
How many symbols should be so employed?
The ASCII committee settled on 4, ASCII characters 28 through 31,
though 8 were proposed~\cite{bemer:18}.
They were called Information Separators~\cite{RFC:183}.
We can do better than that.

There are several issues with having 2 or more Information Separators that
merit careful consideration.

First, consider the sequence {\tt p(q,r;s)}.
$q$ and $r$ are siblings to each other, and descendants of $p$,
and $s$ is 1st cousin to $q$ and $r$.
There are several different more precise meanings this could have.
The semicolon can be given higher precedence than the brackets,
that is, all three of $q$, $r$, and $s$ are children of $p$.
In that case, this particular sequence is invalid,
because $r$ and $s$ cannot be children of $p$ and 1st cousins to
each other.  All children of the same parent must be siblings.

Another interpretation is to allow a single opening bracket to separate a
variable number of generations instead of always one generation.  Then,
since grandchildren
of $p$ can be 1st cousins to one another, all 3 of $q$, $r$, and $s$
must be grandchildren of $p$.  But this idea
has the big disadvantage of adding context dependency to the grammar.
Whether $q$ is a child or a grandchild
of $p$ cannot be known until all the characters between the opening and
closing brackets are scanned.  If a semicolon is found on the same
level as $q$, then $q$ is a
grandchild of $p$.  If there are even deeper separators, $q$ is a
great grandchild or even more distant descendant of $p$.  If none are found,
then $q$ is a child of $p$.

Best is to consider the semicolon as a combined open and close bracket,
{\tt )(}, having the same precedence as any other bracket.
In that case, $s$ is not
a descendant of $p$, $s$ is a nephew of $p$.  That meaning does not
add context. This does
have more invalid strings, for instance the simple sequence {\tt r;s}
is invalid because the brackets are not balanced.

Second, consider how to combine separators with colons.
The colon is a bracket, and should have the same precedence.  Then a
sequence such as {\tt (p:q;r)} means that $p$ is parent to $q$, and not
parent or sibling to $r$.  $p$ is uncle to $r$, $q$ is 1st cousin to $r$,
and $r$'s parent is null.
Perhaps the easiest way to see this
is to reverse the colon transform to get {\tt (p(q;r))}, then
reverse the separator transform to get {\tt (p(q)(r))}.
If $p$ and $r$ are supposed to be siblings, and
$q$ a child of $p$, the correct way to represent that is not to use
semicolon or colon, it is {\tt p(q)r}.

The 2 transforms, colon and separator, are mostly complementary, but
in some cases can compete to reduce the same redundancies.  The following
table shows the results of transforming each of the 14 Dyck words of
length 8 (replacing {\tt ][} with a comma rather than a semicolon,
for greater visual clarity.)

{\tt
\begin{tabular}{rllll}
   & Dyck word &   colon   & separator & both \\
 1 & [[[[]]]] &   [:::]   & [[[[]]]] & [:::]  \\
 2 & [][[[]]] &   [][::]  & [,[[]]]  & [,::]  \\
 3 & [[][[]]] &   [[]::]  & [[,[]]]  & [:,:]  \\
 4 & [[]][[]] &   [:][:]  & [[],[]]  & [[],:]  \\
 5 & [[[][]]] &   [:[]:]  & [[[,]]]  & [::,]  \\
 6 & [[[]][]] &   [[:]:]  & [[[],]]  & [:[],]  \\
 7 & [[[]]][] &   [::][]  & [[[]],]  & [[:],]  \\
 8 & [][[][]] &   [][[]:] & [,[,]]   & [,:,]  \\
 9 & [][[]][] &   [][:][] & [,[],]   & [,[],]  \\
10 & [[][][]] &   [[][]:] & [[,,]]   & [:,,]  \\
11 & [[][]][] &   [[]:][] & [[,],]   & [[,],]  \\
12 & [[]][][] &   [:][][] & [[],,]   & [[],,]  \\
13 & [][][[]] &   [][][:] & [,,[]]   & [,,:]  \\
14 & [][][][] &   [][][][]& [,,,]    & [,,,]  \\
\end{tabular}
}

The last column shows the result of applying the semicolon
transform, followed by the colon transform.  Applied second,
the transform to colon can be blind to the presence of any separators,
and be correct and achieve maximum reduction.  A separator acts as a
bridge, so that a colon can start a list, a natural looking use,
rather than opening the last item of a list.

If the separator transform
is second, then to achieve maximum reduction, as well as a correct
transformation, it has to be done with awareness of colons.
A colon may be opening the last item in a list, and it can be moved
to the head.  {\tt [[]:]} can become {\tt [:,]} by replacing {\tt
]:}, which is an open close pair of brackets, with a separator, and then,
replacing the opening bracket of the previous item in the list with a colon.
This
can be repeated until the colon has migrated to the front of the list.
If the separator transform is done blindly on a sequence with colons,
it can be incorrect.  {\tt[[]][]} is {\tt[:][]} but then replacing
the {\tt ][} with a separator gives
{\tt[:,]}, which is not correct.  Correct is {\tt [[],]}.
Undoing {\tt[:,]} shows that sequence is
actually {\tt [[][]]}, a list of 2 items.

Applying both transforms to the Ackermann function given earlier replaces
a total of 11 bracket pairs with either a single separator (the comma was
used in this example) or a single colon:

\begin{lstlisting}
(define :A x y,
  cond :(= y 0) 0,
        (= x 0, * 2 y),
        (= y 1) 2,
        else :A :- x 1,
                 A x :- y 1)
\end{lstlisting}

Third, what of types?  Should the separated items be the same types?
The traditional meaning of a comma is as a separator only, of untyped
data.

Or, should text adjacent to a comma be interpreted as a type name?
A way to resolve that question is to provide another means to
add a type if desired, and let separators remain separators only.
For example, as mentioned in the section on types, the `\$' character
could be used to indicate an alphanumeric sequence is a type name.  Deeper
separators would need a list of types, or restrictions on elements for which
they can specify a new type, and while notations for that can
of course be invented, there is little point when opening brackets can
accomplish that with reasonable efficiency and without any additional rules.

Fourth, there are different potential meanings for runs of a separator
symbol.  2 adjacent semicolons could mean that there is an empty element
in the middle, like {\tt for(;;i++)} in $C$.
Or, it could mean that the separation between the data elements
on either side is deeper, that is, they are 2nd cousins instead of 1st cousins.
What should 2 semicolons mean, {\tt )()(} or {\tt ))((}?  The former is
the more widely used meaning.  The latter is accomplished by the limited
method of having more Information Separator symbols, which cannot neatly
handle great depths.  It seems useful to have clear and concise ways
to express either meaning.  One way to do this is to have 2 Information
Separators, one for siblings and one for cousins.  The wrinkle is that
repetition of these symbols would have the 2 different meanings.  $n$ of
the sibling separator can mean there are $n-1$ siblings who
have been omitted from the list, while
$n$ of the cousin separator can mean the cousins are $n$th cousins,
being 1st cousins only when $n = 1$.  This approach, combined with an
efficient way to express quantities, discussed next, can express both meanings.

However, another way is not to use the system for expressing quantities,
and then assign different meanings to those quantities, but to use
typing.  A semicolon could be followed by an integer to indicate the
depth of the divide, eg. ``;3'' means the adjacent elements are 3rd
cousins.  Then a run of $n$ semicolons can mean that there are $n-1$
1st cousins in the middle, same as a run of $n$ commas means $n-1$
middle siblings.  This makes it slightly harder to support typed
separators, but of course it can still be done,  That point is moot
if sticking with the traditional meaning of separators being typeless.

A minor matter is that separators have an inherent off-by-one issue.
A comma separated list
usually contains one fewer commas than data items.  Often, specifications
allow a meaningless trailing comma to be present, for the convenience of
programmers.

A big reason to support efficient representation of a cousin
relationship and even reserve symbols especially for it rather than
rely on brackets is that it is a natural way to map multidimensional
arrays to a hierarchical structure.  Another reason is that people are
familiar with and like separators.

\section{Efficient representation of arbitrary quantities}

Infinitely many numbers cannot be represented with single
symbols from a finite set of symbols.

Though we can't collapse arbitrary quantities
to single symbols, we can however do better than using $n$ symbols
to represent $n$, by employing
the same principle used in the Arabic numbering system
that replaced unary numbering systems such as the Roman one and 
hash marks.
All this is well known, as is that a binary
numbering system has the minimum number of symbols needed to represent
quantities of $n$ with $\log n$ symbols.

Can we do even better than $\log n$, represent any arbitrary quantity
of size $n$ with even fewer symbols?  No.
For this question, the Pigeonhole principle applies.
As in data compression, to be able to represent some quantities
of amount $n$ with fewer than $\log n$ symbols (from a finite set of symbols),
other quantities must be represented with more than $\log n$ symbols.
When the amounts are averaged over all quantities $\le n$,
the size is $\log n$, or greater.

Numbering systems can be employed to represent structure.
Rather than come up with more and more symbols to represent greater
and greater quantities, as the ASCII committee did with their
4 separator symbols, can employ 2 symbols in a binary code.

Obviously any one symbol which may be repeated can be made one member of
a set of 2 symbols to be used in a binary encoding.
But if there are many symbols which may be repeated,
finding enough symbols becomes a problem.

Since quantities are so useful, and unused symbols so precious,
a better idea is to reserve 2 symbols for a binary code for quantities only,
for any other symbol that may be repeated.
For example, instead of using 2 kinds of open bracket symbol in a binary
code as in something like {\tt [(([} to represent 9 open brackets,
have {\tt 1001(} mean 9 open brackets, {\tt 1101*} mean 13 asterisks,
and so on.

Still better is to use an escape character and a decimal representation.
The backslash can be used for this, as the only backslash escape sequence
that uses a number is $\backslash0$, to mean the NULL character, ASCII 0.
Then 9 open brackets can be represented with {\tt$\backslash$9(}.
One desirable additional symbol to allow is the minus sign,
for negative quantities.  If only integers are allowed, then there is no
need to overload the meaning of an escaped period for a decimal point
character.

This sort of representation is the well known idea of
run-length encoding (RLE)~\cite{oliver:1952}.  RLE is
simple and easy, even relatively easy for a person to understand without
computer aid.

Of course there is the minor problem that the numeric symbols themselves
cannot be the object of a RLE escape sequence.  There are several easy ways
to resolve that issue.  Easiest is to simply not support repetition of the
digit characters, forcing the use of the traditional method if repetition
of a digit is wanted.  Perhaps next easiest is to employ a terminal symbol.
To keep the representation
one character shorter, the terminal symbol can be optional, used only if
needed to remove ambiguity.

But RLE is very limited in what it can express.  That keeps it dirt simple
and easy to read, but perhaps more expressiveness is desirable, for such
uses as repeating patterns, not only single characters.
One simple example of such
a sequence is the CR/LF pair.  With a trivial amount of
additional syntax, it is possible to efficiently encode repeating patterns.
A further use is as a repetition of an escape.  Suppose one has a string
consisting of many, perhaps over half, of characters that must be escaped.
One traditional method is to inflate by up to double
the quantity of characters by preceding each special character with an escape
character.  That can get difficult for a programmer to read, as seen in
Perl's regular expressions.  A quantity that can be applied to indicate
how many characters
are to be escaped can supersede the traditional escape character method.
This notion is fairly obvious
and has been proposed on a number of occasions, for instance by Rivest
in his draft for S-expressions~\cite{rivest:1997}, 
for what he called ``Verbatim representation'', and with Hollerith constants
in FORTRAN 66~\cite{fortran77}.
Perl's regular expressions has a similar mechanism.

While it is trivial to extend run length encoding to handle repeating patterns,
there are still many other highly redundant strings that this extended
RLE cannot encode, yet are simple to describe.  The question is how
far to go, how much complexity is really useful and can still be easily
encoded?  And, would it still be human readable?

Perhaps an efficient way to represent ``{\tt ))(())((}'' and
larger combinations is also desirable?
To encode such things, more is required.  A simple idea is to support the
encoding of lists of quantities, a vector, rather than a single quantity.  The
escape character can be employed as a separator.  Then what is needed is
agreement on the meanings to assign to the multiple quantities.  For example,
to encode 5 repetitions of a string of length 4, ``$abcd$'', should it be
``$\backslash4\backslash5abcd$'' or
``$\backslash5\backslash4abcd$'' or something else?

But if a vector of quantities is such a good idea, why not a tree of
quantities?  Takes only 2 symbols to represent the structure of a tree.
However, the additional complexity is almost certainly too much to
remain human readable, and there's the question of what uses could we
make of a tree of quantities?

One use for a vector of quantities is for the sizes of the
dimensions of a multidimensional array.  Such a usage creeps into the
domain of external representation of structure.  The interleaving can be
reduced to a single character used as a separator, or removed entirely.
For instance, a 2x3 array with variable sized elements
could be notated as {\tt $\backslash2\backslash3\backslash?$ 1a,1b,1c,2a,2b,2c},
using the same
separator symbol every time, with the division between 1c and 2a known to be
deeper than the rest only because that info was given in the vector of
quantities.  Or that 2x3 array with fixed sized elements could be notated as
{\tt $\backslash2\backslash3\backslash2$ 1a1b1c2a2b2c}.

If means to represent something analogous to a Hollerith constant are provided,
some probably will use it for very complicated objects.  Just serialize the
data, and use the total length of the resulting string as the size.
Supporting runs of the same
symbol, and blocks of data analogous to Hollerith constants, provides enough
for further elaboration if desired, while keeping the notation simpler.

We get away with unary representations, because we stick to
small and simple structures.  If we seldom count higher than 3 or 5,
and almost never higher than 10, tally marks work fine.  A check of
the Firefox source code reveals that only a few programs reached a
nesting depth of 15, with most never exceeding 10, so RLE for opening
brackets and colons, and quantities to indicate the depths of separators
are not going to remove much clutter.
But perhaps flatter structuring has been chosen to avoid
the clutter that would result from deeper nestings.
And block escapes are still a viable use of quantities.

\section{Representing structure with positioning}

The only ASCII control characters still really useful are the 2 for indicating
a new line of text.  Next most used is tab, which has ambiguous meaning
and is easily and often replaced with spaces, except in a few special cases
such as Makefiles.  The rest of the ASCII control characters are very seldom
seen, and when present, modern applications may simply ignore their
meanings~\cite{RFC:7764}.
Of the 132,231 text files in the Firefox 50 source code, just 121 have
ASCII control characters other than the 3 most common: LF, tab, and CR.
A mere 5 files use ANSI escape sequences, which start with the Escape
character (\verb|ctrl-[|, ASCII 27), and that only to set text colors.

ASCII's minimal means of positioning text is sufficient but not efficient or
neat.  One of the
worst inefficiencies is the very repetitive use of spaces to indent lines
of text.  Some ANSI escape sequences address this issue, but not well.
The VT100 series text terminals became popular in large part because they
adopted and extended the ANSI escape sequences.  Yet they have not
been much used outside of terminal control.  They did not grow beyond that
niche to become common within text files.  Colored text is the ANSI escape
sequence most used, yet it is rare.  One of the most common uses of colored
text, highlighting of source code, does not use ANSI at all, even though
editing may still be done in a terminal that supports ANSI.
Rather, text editors parse the
source code being edited to compute which colors to assign, updating
in real time as the user makes changes.  HTML and CSS can specify colors
directly, and are not limited to a tiny 16 color palette.  That and word
processor options have become the way to set text and other colors in
documents.
ASCII and ANSI must use a fixed width font to position text accurately,
and consequently, source code is almost always viewed in such fonts.

What sort of positioning information would be most useful?
Means of clear, easy, and minimal description of position that
best supports useful structures should be leading contenders.
Indentation is the most popular way to express hierarchy through position
alone.  It is so common that even though curly brace languages do not use
indentation, coders are exhorted to use ``proper indentation'' anyway,
so that their source code is more readable.
Perhaps the most prominent and distinctive feature of the Python
programming language is the use of pure positioning to indicate code structure.
Another major use is the alignment of columns, usually for tables.
The ASCII tab character does not do either of these well.

Superscripting and subscripting can be considered a kind of positioning.
It has a major limitation in that it does not scale.
Each successively deeper nesting requires progressively
smaller text, which soon becomes too small to read.

A proposal is to reassign 4 ASCII control characters for indentation.
3 of them can be
increase indent (push), revert indent (pop), and boost indent,
analogous to the 2 brackets
and colon in a closing 3 system. These characters can be invisible and
have a width of zero, not directly affecting the position of text.  They only
change the level of indentation.  The 4th character
can mean forward to the next indentation, replacing the leading spaces
on all indented lines of text.  It could also mean advance to the next line,
but that would be less flexible, wouldn't support situations in which
text such as a line number is wanted before the indentation.

These characters do not specify
the size of an indentation, only the number of levels.  This would
allow individual users to set the indentation size themselves without affecting
others.  It could also make variable width fonts usable, as the problem of
what unit to use to specify indentation sizes is entirely avoided.
It does add one item to the state a text editor or viewer must maintain:
a stack of indentation levels.

Indentation characters could ease editing of source code.
There would be no more need to
shift blocks of code several spaces right or left by changing the number of
leading spaces on each line, whether by manually adding or deleting each
space, or by using some smart editor function such as that assigned to the
tab key in EMACS.

For the columns of tables, need better tab functionality.  It would be
desirable not to rely on the use of a monospace font to recreate the
intended horizontal alignments.  A limitation to dump is any sort
of tiny maximum distance of 8 spaces.  Further, it should be independent
of any indentation setting.  The C1 set of control characters
contains several intended for tabular positioning, but they do not
do enough.  One problem is that the state they set up is global.
Another is that they still implicitly depend upon a fixed font, using the
cursor position for fixing the location of tab stops.
It is basically a copy of the ideas and means used in the most
advanced typewriters, with all their limitations.

The means HTML provides for laying out tables is fairly
comprehensive and flexible, and proven over many years of use.
If better handling of tables is desired in plain text, copying HTML's
handling and a subset of capabilities concerning tables into 
control character functionality seems a good approach.
Lightweight markup languages such as Markdown~\cite{markdown:2016} and
Bulletin Board Code (bbcode)~\cite{bbcode:2011}, arose to satisfy the
desire to be able to create lists and tables in text based forums
more easily than with HTML.
This shows
that many users like text editing to have such capabilities.

\section{Conclusion}

This paper proposed several changes in standard textual notations to eliminate
redundancy that may be hampering the human readability of structured
documents such as source code.  Proving that human readability
is improved was not attempted.  Instead, the paper surmises that some
kinds of redundancy merely add clutter, showed where and how redundancy lurks,
and proposed ways to eliminate it.  Good answers to Lots of
Idiotic Spurious Parentheses have been desired for a long time, and perhaps
until now have not been satisfactory.

Notation that scales and adds expressiveness, and allows much more brevity
without sacrificing clarity, is especially preferred.  Punctuation with
long histories in natural languages, especially English, was tapped as a
guide, in part because
those uses are familiar to people literate in those languages.

The first proposed change was to add a 3rd kind of bracket symbol
roughly equivalent to the meaning of the colon in English, so that a
parent--child relationship represented as ``{\tt(p(c))}''
is instead represented as ``{\tt(p:c)}''.  Proof was given that this 3 symbol
system can collapse all runs of 2 or more closing brackets to a single
closing bracket.

The idea of a universal closing bracket was presented.
''{\tt\{a (b [c] d) e\}}'' can be represented as ``{\tt\{a (b [c] d] e]}'',
reducing the number
of different symbols required, as `{\tt)}' and `{\tt\}}' are no longer needed.

More use of separators was proposed to replace sequences of closing
brackets followed by opening brackets.  ``{\tt((a()b)(c()d))}'' can
be represented with ``{\tt((a,b;c,d))}''.

Ways of adding types to the structure were discussed.

Positioning was recognized as an important way of denoting structure.
It is observed that means of expressing position have been neglected.
Markup languages limit themselves to data, and are not
much used for writing of other structured information such as source code.
Moreover, by using visible text to express position, and requiring
translation with special tools such as a web browser, they fail at the goal
of using position alone to express structure.
The means provided in ASCII work only with
monospace fonts, and require much wasteful redundancy.  Repurposing
some of the unused control characters to better support indentation and
tabular structure was proposed. 

Together, these changes reduced the number and quantity of symbols
needed.  They improved the amount of data compression obtained by general
purpose data compression programs.  They reduced the size of the source code.
Whether the goal of greater human readability was also achieved was not
studied, but it was surmised that removing redundancies in the notation
does help with readability.

\end{document}